# From Public Data to Private Information:
# The Case of the Supermarket

Vincent C. Müller

*Anatolia College/ACT*
*P.O. Box 21021*
*Pylaia, Greece*
*+30-2310-398 211*
*http://www.sophia.de*
*vmueller@act.edu*

**Abstract:** The background to this paper is that in our world of massively increasing personal digital data any control over the *data about me* seems illusionary – informational privacy seems a lost cause. On the other hand, the production of this digital data seems a necessary component of our present life in the industrialized world. A framework for a resolution of this apparent dilemma is provided if by the distinction between (meaningless) *data* and (meaningful) *information*. I argue that computational data processing is necessary for many present-day processes and not a breach of privacy, while collection and processing of private information is often not necessary and a breach of privacy. The problem and the sketch of its solution are illustrated in a case-study: supermarket customer cards.

1. **Privacy and private information in the digital world:
   A framework**

   1.1. **The general problem:
        Privacy in the digital world is a lost cause**

The concerns about informational privacy have become more urgent with the advent of widespread digital computer technology. This concerns two types of activities. First, matters that I, as an individual, now do with computers and that were normally done with other media some 25 years ago (e.g. word processing, mail, telephone, photographs & videos, reading news, address book, diary, bibliography, music, etc. etc.). Apart from these matters that are obvious to me, since I myself do them on a digital system, a very large amount of data is produced by other non-computational activities of everyday life, e.g. driving (car tracking [GPS, navigation systems, toll systems], road tolls, video surveillance, insurance data, police data, …), speaking on the telephone (connection data, cost data, location, content storage, …), taking photographs (image, time & date, technical data, location, …) or just buying things (credit card data, bank data, customer data, location, …) – not to mention the data produced by as yet uncommon applications, such as 'ambient intelligence', remote health services, etc. These developments are clearly increasing: more systems become integrated (e.g. telephone-computer-camera-music player) and more systems become computerized. In addition to this production of data, the possibilities of analyzing and storing them forever have greatly increased.

Let me make it quite clear that I do not generally deplore these developments: I am quite glad that I can now learn of an interesting academic conference, get further information, register, read literature, send an abstract, receive reviews, communicate with colleagues, write a formatted text, revise it, etc.; all without leaving my desk, at negligible cost and with no delays. Of course, some of the developments sketched above are driven by profit interests and political interests that I happen not to approve of (e.g. state surveillance), but it is equally evident that a lot of these are in my personal interest and in that of others.

On the other hand, this digitalization means that a lot of information about me is now accessible to other people and that I have lost control of this information. In other words, the developments bring with them a loss of *informational privacy*. I use this term in the classic sense: "Privacy is the claim of individuals, groups or institutions to determine for themselves when, how and to what extent information about them is communicated to others." (Alan F. Westin 1967, 7, quoted from Rosenberg, 2004, 349).

This is where the general dilemma lies: I want privacy and I want a digital life. It seems that I can't have both and it seems that I can't even reject the latter. In other words, it seems that privacy is a lost cause in the digital world.

### 1.2. The framework

In this paper I wish to sketch a general framework that could provide at least a partial way out of this dilemma. The framework relies with a philosophical distinction and I will then investigate in a case-study whether that framework holds water in practice.

A digital computer system operates according to algorithms on tokens of types that are in specific digital states and produces further tokens of these types (for details on digital states, see Müller, 2008). In current systems these are typically just two states; the systems are binary. These basic states are used to represent higher level states, e.g. numbers, letters or truth-values, and these higher level states are again used to represent higher level properties, e.g. bank account balance, a beep, or an error message. So, if I send an SMS message saying "I will be back at 5.", this message is in digital states on several levels and at its destination it will hopefully produce a particular pixel image on a certain mobile phone screen. Is there a privacy problem here? If the message is just a sequence of binary data that is algorithmically processed by several conventional computing machines, then this message has no *meaning* for these machines, nothing is understood. If there is no meaning, then no information about me is conveyed and my privacy is wholly unaffected by the process. Our right to privacy is geared to human relations, i.e. it concerns what another human (or at least intelligent agent) can understand, and nothing of the sort is taking place here. Contrast this computational processing with the case of a Morse telegraph where human operators on both ends must encode and decode the message – normally understanding it in the process.

Generally, I think we must distinguish between *data*, meaningless digital states that are processed, and *information*, which is meaningful for someone. The qualification 'for someone' is needed to capture cases where something is data for one person but information for another – if I look at a road sign written in Georgian script, this is just data for me, for a Georgian speaker it is information. (Incidentally, in this and many other cases, it might not even be possible to distinguish the digital data, e.g. to copy the sign in such a way that it would distinguish the letters properly.)

If this is the right framework, then privacy is only an issue if persons come into play, at least at some point. This point might be after a lot of digital data processing, e.g. if the NSA processes my SMS as part of data mining for counterterrorism *surveillance* – an issue I investigated in (Müller, 2009). Given that the data is turned into personal information at some point, a privacy issue does arise in these cases.

Quite simply, mere *data* becomes *information* when it has meaning to someone. This can happen to personal data that I would like to keep private (in surveillance), but it can also happen to data that I control only to some extent, or even not at all. What happens when I buy something at the local supermarket is data that I do not control, so I will investigate the general framework of data vs. information in the case study of supermarket customer cards.

## 2. Case study: the local supermarket

### 2.1. What supermarkets do for marketing

Supermarket "customer cards" or "loyalty cards" are given away to customers with a promise of participation in special rebates or cash-returns to loyal customers. Of course, no guarantee is given that the owners of the card actually benefit. In fact, some studies have found that rebates were typically *lower* than before the introduction of the cards (Albrecht, 2001, 536). Crucially, customers are not informed – and often not aware – that their shopping data is collected.

It is apparently far from clear that these schemes do much work in terms of customer loyalty (McIlroy and Barnett, 2000), especially in areas like supermarket shopping where customers are not strongly related to a particular business and thus 'relationship marketing' is likely to be less successful (Boedeker, 1995). So, the actual point of the 'loyalty' card is less customer loyalty than data collection. This point the customer is typically not informed about.

The supermarket will store two kinds of data, and link them: data on the *customer* and data on the *purchases*.

The card stores identifying data, as far as this is accessible to the supermarket. This will typically include full name and address, often a telephone number and e-mail address, the date of birth and gender. This set of personal data is sometimes supplemented by financial data, if the card is combined with credit or debit card functions. The card may also be combined with further uses, e.g. insurance coverage or employee benefits, that imply further personal data. In practice, any kind of card can be used that identifies the customer, so a credit card or even an ID card would do. The more certain the identification, the better the integration with other data.

At the moment of purchase and with the help of the card, the supermarket will store the customer ID, transaction date & time, item bought (with number of items, classifications, price levels, promotions, relations to other items, etc. etc.), means of payment and other information that it finds relevant. It may combine that information with other data, such as personal movement through the store or even the fate of purchased items outside it (e.g. through RFID).

What the supermarket is trying to achieve is something that other businesses already have, by design. If sales take place over electronic devices, e.g. web sites then customers are (at least partially) identifiable through customer numbers, credit cards, 'cookies' or static IP numbers, and thus the business already has all that data on customers and purchases available. This is standard practice on the Internet. (If you have any doubts, check the cookies in your web browser.)

Apart from being used, this data is also stored indefinitely, for uses still to be determined. This work is normally outsourced to specialized IT firms, who also sell the data to other interested parties.

## 2.2. A sketch of the problem:
## Data mining on customer and purchase data

Imagine that there was a person at your neighborhood store who noted down the name of each customer and what they bought and when. I presume you would not be happy. – This is precisely what the customer card does, at least on the level of data. Here is an example of resulting transaction data, depicted in a binary matrix (Hand, Mannila and Smyth, 2001,8):

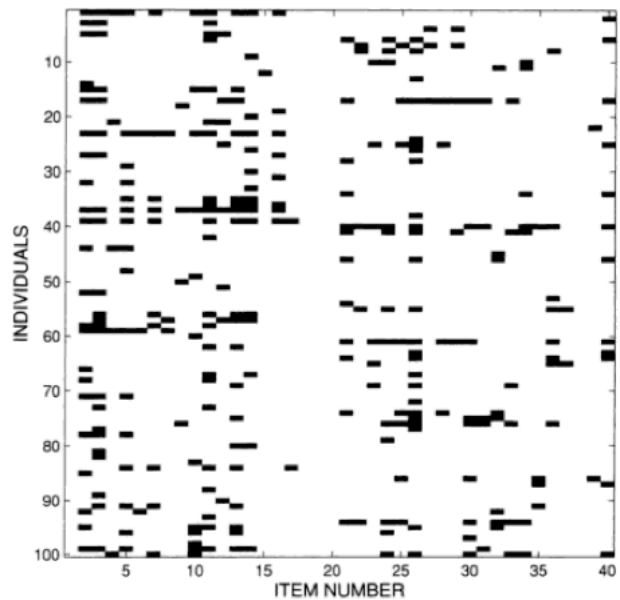

Figure 1.1 A portion of a retail transaction data set displayed as a binary image, with 100 individual customers (rows) and 40 categories of items (columns).

A typical response to this kind of activity from privacy concerned academics is this very useful study: "The shift from a paper-based to an electronic-based society has dramatically reduced the cost of collecting, storing and processing individuals' personal information. As a result, it is becoming more common for businesses to 'profile' individuals in order to present more personalized offers as part of their business strategy. While such profiles can be helpful and improve efficiency, they can also govern opaque decisions about an individual's access to services such as credit or an employment position. In many cases, profiling of personal data is done without the consent of the target individual." (Camenisch, Sommer, Fischer-Hübner et al., 2005, 20)

Having said that, 'profiling' particular individuals is not normally the point of customer cards, but rather the support of general and particular business decisions about store supplies, store location, product presentation, pricing, etc. For these purposes, the data sets produced must be subject to analysis, typically using techniques of data mining. "Data mining is the discovery of interesting, unexpected or valuable structures in large datasets" (Hand, 2007, 621) in particular, it is used in data that does not yet provide the structure that one is looking for. It looks roughly for two kinds of things:

A global *model* in data mining is a statement that applies to all the data, thus says something about each data set. A global model does not turn any particular data into information but says something about all data – e.g. whoever bought item 2 had a high probability of also buying item 3.

What is more interesting for our purposes is what is called a local *pattern* where a specific set of data is singled out because of 'interesting characteristics' (this technique will also produce "false negatives" and "false positives"). One aim of pattern data mining is to find; "association rules" (who buys one type of items also buys another type), the classical algorithm for which is described in (Jong Soo, Ming-Syan and Philip, 1995). These 'patterns' can help the business decide which items to stock in which store and how, it can help planning for seasonal items, etc. etc.

A problem that remains with this technique is that it only finds correlations, not causal relations. It is thus never certain that a certain correlation is not accidental – though the probability of such findings can be minimized by large and diverse enough data sets that offer enough repetitions (rather than one-off events) (Hand, Mannila et al., 2001, esp. 119). Many technical papers deal with the details of handling these problems in the supermarket environment, e.g. (Lawrence, Almasi, Kotlyar et al., 2001).

### 2.3. Who's data? Who's information?

This data is property of the supermarket, it can and is thus sold if this provides a profit. (Some of the data is anonymized before the sale, which may or may not assure its anonymity.) It may also fall into other hands, e.g. lawyers, law enforcement, secret agencies, etc. The data is in the hands of supermarket employees, not especially protected by state agencies or procedures, so misuse and errors are certain to occur. What can be done with this data is unlimited. For example, the UK Internal Revenue Service has demanded access to customer data in order to verify income statements in tax records (Albrecht, 2001, 539).

The urgency of our problem can be underlined by the combination of two remarks: "'You are your information', so anything done to your information is done to you, not to your belongings." (Floridi, 2006, 111) and the old saying "Tell me what you buy and I tell you who you are!"

Having said that, is there any personal information in the matrix depicted above? No, we are just given a set of unidentified individuals and a set of unidentified items. This is just data. Presumably it is necessary for the purposes of the supermarket to identify the items, so the system will contain a list which maps the numbers to products for sale. Still, as long as the individuals are not identified, this is not personal information. However, if there is a database of all customer cards, then the supermarket has a matching tool for combining this information in such a way that it does become personal information (Mr Smith buys a lot of alcohol, etc.). Thus, if there is such a database – and in all schemes I know of there is – then common privacy concerns apply, i.e. individuals must be informed and asked for consent to the spread of this information. In this case, the collection of personal information does not even seem particularly useful, except that it helps the supermarket find out the location of customers (for store location). All other aims can be reached with an anonymous card that give the benefits to the person who happens to present it. The collection of data is useful, the collection of information is superfluous.

## 3. Conclusion: The right to information, not data

The increasingly digital world poses serious challenges with its production of large amounts of personal digital data. However, there are technical and political means to reduce the accumulation of this data, and, more importantly, to prevent the turning of this data into personal information. This distinction between data and information has implicitly been used already in many cases.

The little case study of supermarket customer cards has shown that this distinction can be made in practice and that it can provide for a means to achieve the aims of the data-collector while respecting the right to privacy of the customers. This suggests that political work should go into the direction of turning information into data, rather than just fighting data collection per se (which remains an important aim, however). Philosophically and practically, we will need to sharpen the distinction in such a way that it can cover the many diverse cases and needs.